 \documentclass[smallabstract,smallcaptions]{dccpaper}

\usepackage{epsfig}
\usepackage{citesort}
\usepackage{amsmath}
\usepackage{amssymb}
\usepackage{color}
\usepackage{url}
\usepackage{subfigure}
\usepackage[linesnumbered,ruled,noend,vlined, scleft,nofillcomment]{algorithm2e}

\newlength{\figurewidth}
\newlength{\smallfigurewidth}

\begin{document}

\title
{\large
\textbf{Bitvectors with runs and the successor/predecessor problem\thanks{\scriptsize{Funded in part by European Union’s Horizon 2020  Marie Sk{\l}odowska-Curie grant agreement No 690941; MINECO  (PGE and FEDER) [TIN2016-78011-C4-1-R]; Xunta de Galicia (co-founded with FEDER) [ED431G/01; ED431C 2017/58; IN852A 2018/14], Xunta de Galicia (co-founded with GAIN) [IN848D-2017-2350417] and FPU Program [FPU16/02914]}}}
}

\author{%
 Adri\'an G\'omez-Brand\'on\\[0.5em]
{\small\begin{minipage}{\linewidth}\begin{center}
\begin{tabular}{c}
 Universidade da Coru\~na \\
 Laboratorio de Bases de Datos, CITIC \\
A Coru\~na, Spain\\
\url{adrian.gbrandon@udc.es} \\
\end{tabular}
\end{center}\end{minipage}}
}

\maketitle
\thispagestyle{empty}

\begin{abstract}
	
The successor and predecessor problem consists of obtaining the closest value in a set of integers, greater/smaller than a given value. This problem has interesting applications, like the intersection of inverted lists.  It can be easily modeled by using a bitvector of size $n$ and its operations \textit{rank} and \textit{select}. However, there is a practical approach \cite{Nav16}, which keeps the best theoretical bounds, and allows to solve \textit{successor} and \textit{predecessor} more efficiently. Based on that technique, we designed a novel compact data structure for bitvectors with $k$ runs that achieves \textit{access}, \textit{rank}, and \textit{successor}/\textit{predecessor} in $O(1)$ time by consuming space $O(\sqrt{kn})$ bits. In practice, it obtains a compression ratio of $0.04\%-26.33\%$ when the runs are larger than $100$, and becomes the fastest technique, which considers compressibility, in \textit{successor}/\textit{predecessor} queries. Besides, we present a recursive variant of our structure, which tends to $O(k)$ bits and takes $O(\log \frac{n}{k})$ time.

\end{abstract}

\Section{Introduction}

One of the main computational tasks in a search engine is to look for those documents that contain a set of words. In order to speed up that search, those engines use inverted lists. Each inverted list corresponds to a word and stores as an increasing sequence the document identifiers of the documents where that word occurs. Most of the time, the query received by a search engine carries more than one word, to know where all the words appear together,  the search engine needs to intersect various inverted lists. The optimal intersection of two lists can be easily solved by iterating over both of them in alternate form \cite{culpepper2010efficient}. In each iteration, the search engine looks for a value in the second list, $v_2$, equal to or higher than the value from the first list, $v_1$. If they are identical, $v_1$ is part of the solution and iterates to the next value in the first list. Otherwise, the iterator of the first list skips those values lower than $v_2$. Therefore, it needs an efficient mechanism that can find an equal or higher value in the other list, which is known as the successor problem. 

Let us formalize the successor and predecessor problem, considering a set of integers $S = \{x_1 < x_2 < \dots < x_m\}$, the successor ($succ(x)=x_i$) of a given value $x$ returns the minimum value $x_i \ge x$ of $S$. Analogously, the predecessor of $x$ ($pred(x)=x_i$) returns the maximum value $x_i \le x$ of $S$. Assuming $n=x_m$ and $m=|S|$, both problems can be modeled by using a bitvector $B[1, n]$ which contains $m$ 1s located at positions $x_i$ for all $1 \leq i \leq m$, and solved in $O(1)$ time with the two classical operations on bitvectors: \textit{rank} and \textit{select} \cite{jacobson1988succinct,munro1996tables,clarktrees,gonzalez2005practical}.

In some scenarios, the bitvector $B$ can contain the set bits clustered together in $k$ runs; hence $B$ contains $k$ runs of 1s and $k \pm 1$ runs of 0s. There is a structure, \textit{oz-vector} \cite{Nav16}, which compresses the bitvectors exploiting its runs. The \textit{oz-vector} transforms the input bitvector into two sparse bitvectors \textit{O} and \textit{Z}, which mark the lengths of the runs of 1s and 0s, respectively. Since those bitvectors are sparse, they are very compressible, and the \textit{oz-vector} can obtain good compression ratios in practice. However, for solving \textit{succ} and \textit{pred}, it requires $O(\log k)$ time.
This study aims to propose a new structure, \textit{zombit-vector}, which compresses bitvectors with runs and supports $succ$ and $pred$ in $O(1)$ time. \textit{zombit-vector} splits the bitvector into blocks of fixed size $\beta$ and classifies them into three sets of blocks depending on their information: full of 0s ($\mathcal{Z}$), full of 1s ($\mathcal{O}$), and containing 1s and 0s ($\mathcal{M}$). Representing this information needs $O(k\beta + \frac{n}{\beta}) + o(n)$ bits. As the optimal value of $\beta$ is $\sqrt{\frac{n}{k}}$, it can solve \textit{access}, \textit{rank}, \textit{succ}, and \textit{pred} in $O(1)$ time with $O(\sqrt{kn})$ bits. Besides, we present a variant which constructs a \textit{zombit-vector} recursively. That recursive technique converges to $O(k)$ bits and can solve those operations in $O(\log \frac{n}{k})$.

We compare our proposal with different compact data structures for bitvectors that can solve \textit{succ} and \textit{pred} efficiently. In the experimental evaluation, we can observe that \textit{zombit-vector} obtains the best response times in all settings. It becomes $5-12$ times faster than our immediate competitor and occupies $0.04\%-26.33\%$ of the plain bitvector, when the mean length of runs is higher than 100.

\Section{Background}
A bitvector $B[1, n]$ is an array of bits whose size is $|B|=n$, and each position can acquire the two possible bit values, 0 and 1. Usually, it is used to represent the values of $S = \{x_1, x_2, \dots x_m\}$, a subset of an universe $\{1, 2, \dots n\}$, by setting $B[x_i]=1$ such that $x_i \in S$. Mainly, they support two operations: $rank_\alpha(B,x)$, which returns the number of bits set to $\alpha$ within the interval $B[1, x]$; and $select_\alpha(B,x)$, which returns the position of the $x$-th $\alpha$ value in $B$. In theory, both operations can be solved in $O(1)$ time by using additional structures which require $o(n)$ bits of extra-space \cite{jacobson1988succinct,munro1996tables,clarktrees,gonzalez2005practical}. 

Therefore, successor and predecessor problem can be modeled by setting $B[x_i]=1$ where $x_i \in S$ and both operations, with respect to a position $x$, can be solved as $succ(B,x)=select_1(B, rank_1(B,x-1)+1)$ and $pred(B,x)=select_1(B, rank_1(B,x))$. Though both are $O(1)$ time, there is a more practical structure for \textit{succ}/\textit{pred} \cite{Nav16}, which keeps the $o(n)$ extra-space and $O(1)$ time. In practice, it achieves less space and better response times than using \textit{rank}, and then \textit{select}. That structure is similar to the classical rank structure \cite{jacobson1988succinct,munro1996tables,clarktrees,gonzalez2005practical}, but instead of storing the number of ones preceding a position, it stores the location of the next/previous 1-bit. Since predecessor and successor are symmetrical, from this point on, we only refer to successor.

\SubSection{Zero-order entropy}
Notice that, in a plain bitvector, we are using $n$ bits of space, which is the worst-case optimal, and achieves $rank$, $select$, and $succ$  in $O(1)$ time with additional $o(n)$ space. However, a better lower bound of the representation of $S$ is $\mathcal{B}(n, m)=\lceil \log\binom{n}{m}\rceil$. In order to improve the worst-case optimal, in~\cite{raman2002succinct,pagh1999low} they propose techniques, which get $O(1)$ time in \textit{rank} and \textit{select} by using $\mathcal{B}(n, m) + o(n)$ bits.  Consequently,  $succ$ can be solved in $O(1)$ time. The space result is approximately $nH_0$,  where $H_0$ is the zero-order entropy of $B$. Some studies show new lower-bounds \cite{golynski2007optimal,miltersen2005lower} and confirm that the space of \cite{raman2002succinct,pagh1999low}  is almost the optimal. 

\SubSection{Sparse bitvectors}

Those bitvectors where $m \ll n$ are well-known as \textit{sparse bitvectors}. 
In sparse bitvectors the extra $o(n)$ space can be huge, for this reason there are some proposals, like \textit{rec-rank} and \textit{sd-array} \cite{okanohara2007practical}, focused on avoiding that dependency. 

The first one splits $B$ into  blocks of a given fixed size and classifies those partitions into two types: $Z$, full of 0s, and $NZ$, the block contains at least one 1-bit. The kind of each block is stored into a \textit{contracted} bitvector, $B_c$, and the $NZ$ blocks are grouped together by concatenating them preserving the order into an \textit{extracted} bitvector, $B_e$. This process is repeated recursively over $B_e$ until it is not sparse. In total, it takes $\log\frac{n}{m} + m+ o(n)$ bits and can solve \textit{rank}, \textit{select}, and $succ$ in $O(\log\frac{n}{m})$ time.

The \textit{sd-array} defines a parameter $r=\lfloor \log \frac{n}{m} \rfloor$ and each value $S[i]$ is divided into the $r$ lowest bits ($l_i$) and the $\lceil \log n \rceil - r$ most significant bits $(h_i)$. Notice that each $h_i$ covers an interval of values $[h_i\times2^r, (h_i+1)\times2^r)$. With this information, the \textit{sd-array} builds two elements, $L$ and $H$.  $L$ is an array composed by each $l_i$, and $H$ is a bitvector that indicates in unary how many elements of $L$ are covered by all possible $h_i$. Since $L$ can be stored by using $m\log\frac{n}{m}$ bits, $H$ uses at most $3m$ bits, and an additional structure of \textit{select}; the required space is $m\log\frac{n}{m} + O(m)$ bits. It is able to solve \textit{select} in $O(1)$ time, but \textit{rank} and $succ$ operations take $O(\log\frac{n}{m})$.

\SubSection{Bitvectors with runs}

Occasionally, a bitvector has the same distribution of 1s and 0s; however, both bits are clustered together forming runs. Since the number of 1s and 0s are similar, the zero-order entropy cannot capture the compressibility of bitvectors with runs. Therefore, the previous techniques are not useful in this case.

Nevertheless, in \cite{Nav16} they propose a structure called \textit{oz-vector}, which makes possible to use those previous techniques by transforming $B$ into two sparse bitvectors ($O$ and $Z$). The transformation consists in computing the length $\ell$ of every run of 1s and 0s, and storing those lengths with unary code ($10^\ell$) in $O$ and $Z$, respectively. Consequently, the length of the first run of 1s is the distance between the first and second 1-bit in $O$. By using the \textit{sd-array} in $O$ and $Z$, the \textit{oz-vector} uses $k\log\frac{2n}{k} + O(k)$ bits and solves \textit{select} in $O(min(\log k, \log\frac{n}{k}))$ and \textit{rank} in $O(\log k)$. Therefore, $succ$ can be solved in $O(\log k)$.

\SubSection{Hybrid bitvectors}
As it is shown in the previous scenarios, we can find different types of bitvectors depending on its number of 1s or how clustered are those bits. In \cite{karkkainen2014hybrid}, the authors propose a structure (\textit{hybrid-vector}) that can adapt its compression technique according to the features of its input.  The \textit{hybrid-vector} splits the input into blocks of a fixed size $b$, and each one is encoded individually.  There are three possible types of block encoding: (i) minority bit positions, which stores only the positions of all 1s or 0s, the one which has fewer occurrences; (ii)  run-length encoding, stores the length of every run; and (iii) plain, in case that the previous options are not satisfactory, the block is encoded in plain form. Over those blocks, in order to speed up \textit{rank} and \textit{select}, there is an auxiliary structure of super-blocks storing the accumulative rank. In the worst case it takes $n+o(n)$, but if it has $k$ runs and $m$ minority ones, it only uses $min(k,m)\lceil\log b \rceil +o(n)$ bits. With this space, it achieves \textit{rank} in $O(1)$, but \textit{select} and \textit{succ} require $O(\log n)$ time.

\Section{zombit-vector}

Our proposal, \textit{zombit-vector} is designed to compress bitvectors with runs and solve successor and predecessor operations in $O(1)$ time. The main idea is to divide the input into blocks in such a way that most of the blocks are uniform (all 0s or all 1s). With this approach, our structure only needs to store the information contained by non-uniform blocks. 
%Additionally, the compression can be improved by applying this technique recursively. 

\begin{figure}[t]
	\centering    
	\includegraphics[width=0.9\textwidth]{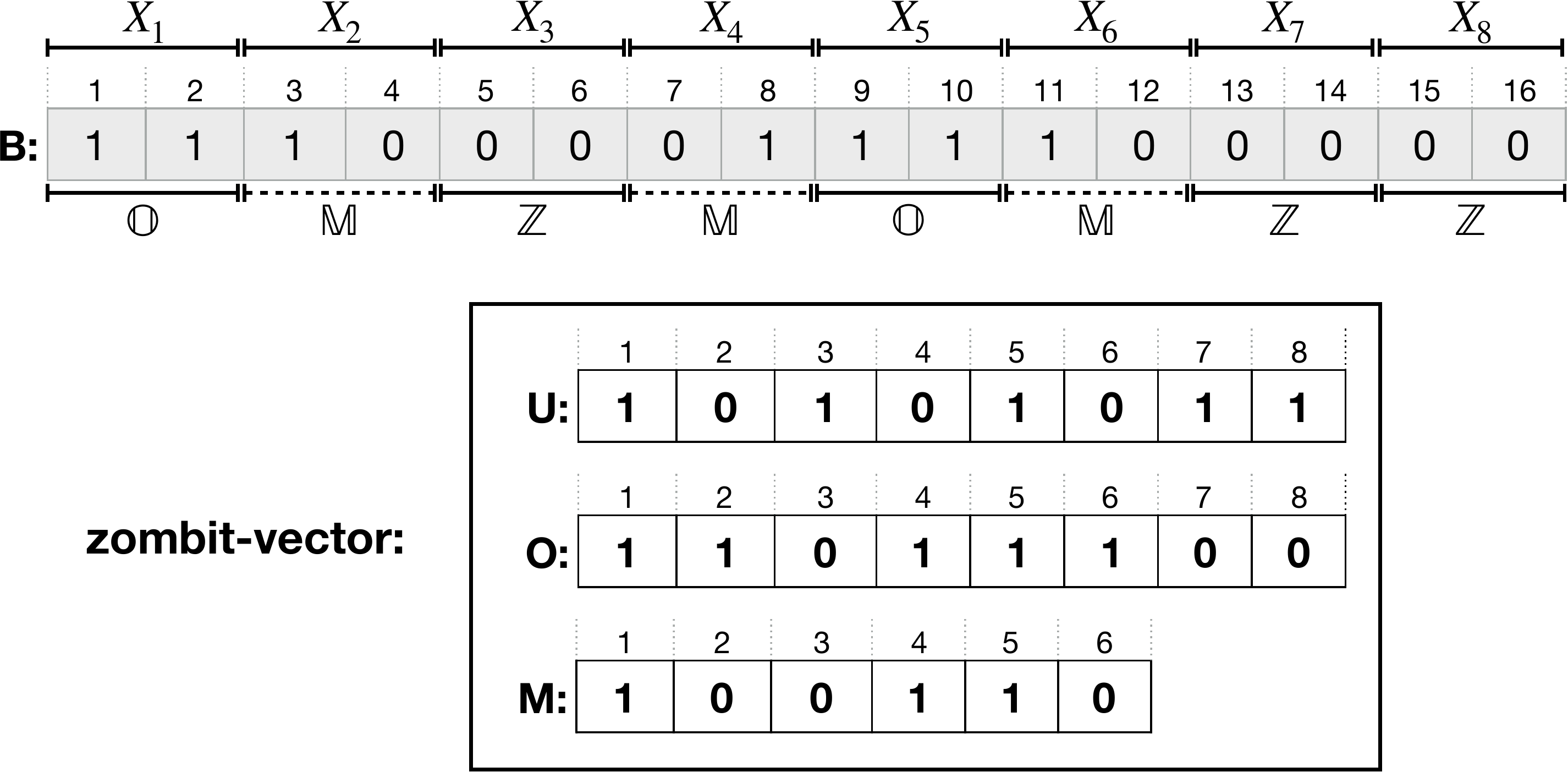}
	\caption{Example of \textit{zombit-vector} with $\beta=2$.}
	\label{fig:zombit-vector}
\end{figure}

\SubSection{Structure}

Given a bitvector $B$ of size $|B|=n$ with $k$ runs of 1s and $k \pm 1$ runs of 0s, \textit{zombit-vector} defines a size of block $\beta$ which splits $B$ into $\lceil \frac{n}{\beta} \rceil$ partitions, obtaining a set of blocks $\{ X_1, X_2, \dots, X_{\lceil \frac{n}{\beta} \rceil} \}$. Each block $X_i$ can be classified  into three different sets of blocks depending on its data: uniform blocks full of 0s ($\mathcal{Z}$), uniform blocks full of 1s ($\mathcal{O}$), and mixed blocks, those which contain both bits ($\mathcal{M}$). As a consequence, the structure contains $u=|\{\mathcal{Z} \cup \mathcal{O}\}|$ uniform and $t=\lceil \frac{n}{\beta} \rceil - u$ mixed blocks. The classification of each block can be represented by using two plain bitvectors: $U$ and $O$. The bitvector $U[1, \lceil \frac{n}{\beta} \rceil ]$ marks which $X_i$ is uniform by setting $U[i]=1$ when $X_i \in \{\mathcal{Z} \cup \mathcal{O}\}$.
Then, we use the bitvector $O[1, \lceil \frac{n}{\beta} \rceil ]$ to represent which block contains at least one 1-bit, it means $O[i]=1$ when $X_i \in \{\mathcal{O} \cup \mathcal{M}\}$. 
Additionally to this classification, we need to store the data of every mixed block. For that purpose, we use a bitvector $M[1, t\times\beta]$ which appends the information of each mixed block together, preserving the order in $B$. In total, we need $O(k\beta + \frac{n}{\beta})$ bits, and since the optimal $\beta=\sqrt{\frac{n}{k}}$, we can reduce the space to $O(\sqrt{kn})$ bits.

Figure~\ref{fig:zombit-vector} shows an example of \textit{zombit-vector} built over the bitvector $B$ with $\beta=2$. On top of Figure~\ref{fig:zombit-vector}, $B$ is divided into 8 blocks $\{X_1, X_2, \dots, X_8 \}$. 
On the bottom of each block, we can observe the set which contains such block and solid or dashed lines, which distinguish uniform and mixed blocks, respectively. For instance, $X_1 \in \mathcal{O}$ is uniform, thus it gets a solid line, but $X_2 \in \mathcal{M}$ is mixed, and it is represented with a dashed line. As $X_2, X_4, X_6$ are the mixed blocks, in the bitvector $U$ of the \textit{zombit-vector} structure, all the bits are set to 1 except $2, 4$ and $6$. Concerning the $O$ bitvector, it has only 0-bits at positions $3, 7$, and $8$ because $X_3, X_7, X_8$ belong to $\mathcal{Z}$. Finally, the mixed blocks are sequentially stored in the bitvector $M$, hence $M[1,2]$ corresponds with $X_2$, $M[3,4]$ with $X_4$ and $M[5,6]$ with $X_6$.

\SubSection{Basic operators}
For a better explanation of the operations that can be solved in the \textit{zombit-vector} structure, let us define two basic operators:
\begin{itemize}
	\item $beg_i$ and $end_i$: given a position $i$, which represents the $i$-th block, it returns the first/last position, respectively, where the information of the $i$-th block is located. The result can be computed as $beg_i=(i-1)\times\beta+1$ and $end_i=i\times\beta$. 
	\item $\Delta_i$: computes the distance of $i$ with respect to the first position of its block as $(i-1)\mod\beta$. For example in Figure~\ref{fig:zombit-vector}, $\Delta_8=1$ because of $\beta = 2$ the first position of its block is $7$, hence the distance is $8-7=1$.
\end{itemize}

\SubSection{Access}
Access operation retrieves the value at a given position $i$. That information is stored on a block $X_j$, where $j=\lceil \frac{i}{\beta}\rceil$. If $X_j$ is uniform ($U[j]=1$), all the values inside that block are identical. Since  $O[j]$ indicates when a block contains at least 1-bit, the uniform block is full of 1s when $O[j]=1$; otherwise, it is empty. Therefore, in the uniform case, access can be solved by returning $O[j]$. For instance, with $i=5$ the solution is in $X_3$, an uniform block ($U[3]=1$), thus the algorithm returns $O[3]=0$.

Otherwise, when $X_j$ is mixed ($U[j]=0$), the number of mixed blocks up to $j$ is computed as $q = rank_0(U, j)$. Consequently, the value of $i$ is inside the $q$-th mixed block, specifically, $\Delta_i$ positions after the first position of that block. As the data of the $q$-th mixed block are at $M[beg_q, end_q]$, the value corresponds with $M[beg_q+\Delta_i]$.
For example, in Figure~\ref{fig:zombit-vector} to retrieve $B[7]$, the algorithm computes $j=4$ and checks $U[4]$. $X_4$ is a mixed block because $U[4]=0$ , in particular, it is the second mixed block ($q=rank_0(U,4)=2$). Hence, the algorithm returns $M[beg_2+\Delta_7]=M[3+0]=0$. As we can observe, in both cases, this operation can be solved in $O(1)$ time by using $o(n)$ bits of additional space for \textit{rank} in plain bitvectors.

\SubSection{Rank}
$rank_1(B,i)$ computes the number of 1s in $B$ up to a position $i$ which belongs to a block $X_j$. In \textit{zombit-vector}, the algorithm starts computing the number of 1s preceding to $X_j$. Let us define $p'/q'$, as the number of uniform/mixed blocks previous to $X_j$.\footnote{Note that $p'=rank_1(U, j-1)$ and $q'=rank_0(U, j-1)$.}  Therefore, the uniform blocks include $b_1=p'-rank_0(O, j-1)$ blocks full of 1s. Since there are $b_1$ blocks with $\beta$ 1s each, and $q'$ mixed blocks that accumulate $rank_1(M, end_{q'})$ 1s, the number of 1s before $X_j$ is $r=b_1\times\beta + rank_1(M, end_{q'})$.

After computing the number of 1s previous to $X_j$, the algorithm updates $r$ depending on the number of 1s in $X_j$. If $X_j \in \mathcal{Z}$, the block is absent of 1s and the algorithm returns $r$. When $X_j \in \mathcal{O}$, the number of ones contained by $X_j$ up to $i$ must be considered. Since $X_j$ is full of 1s, thus it contains $\Delta_i+1$ extra 1s, the solution is $r + \Delta_i+1$.
 Otherwise, $X_j $ belongs to $\mathcal{M}$, which is the $(q'+1)$-th mixed block. The number of ones up to $i$ inside $X_j$ is computed as $b=rank_1(M, beg_{(q'+1)} + \Delta_i) - rank_1(M, end_{q'}) $, thus the solution becomes $r+b$. Notice that, we can simplify that formula as $b_1\times\beta +rank_1(M,beg_{(q'+1)}+\Delta_i)$, the number of 1s previous to $X_j$ in those blocks full of 1s, and the number of 1s inside the mixed blocks up to the queried position. For example, in Figure~\ref{fig:zombit-vector} for solving $rank_1(B, 8)$ where $j=4$, the algorithm gets $p'=2$, $q'=1$, and $b_1=1$.  $X_4 $ is a mixed block, hence the algorithm computes the first addend  as $b_1\times\beta=2$, and the second one as $rank_1(M,beg_{(1+1)}+\Delta_8)=rank_1(M, 3+1)=2$. The addition of these two values is the solution, $2+2=4$. As we can observe, the \textit{rank} operation on \textit{zombit-vector} only requires \textit{rank} on plain bitvectors, which can be solved in $O(1)$ time by storing $o(n)$ bits of extra-space.  It is important to notice that we do not discuss $rank_0(B,i)$ because it can be solved as $i - rank_1(B,i)$.
 
\SubSection{Successor}

Given a bitvector $B$ and a position $i$, $succ(B,i)$ returns the lowest index which contains a 1-bit at $B[i, n]$. For example, in Figure~\ref{fig:zombit-vector}  $succ(B,3)=3$ and $succ(B,6)=8$. This operation can be solved in the \textit{zombit-vector} following Algorithm~\ref{alg:successor}.   Firstly, at Lines 1-2, the algorithm computes the block that contains $i$ ($X_j$), and stores the number of mixed blocks up to $X_j$ into $q$. There are two cases where  $X_j$ includes the solution. The first case occurs when $X_j \in \mathcal{O}$, at Line 5. Since $X_j$ is full of 1s the successor is the current position $i$. The second one happens when $X_j \in \mathcal{M}$ and the next 1-bit is inside $X_j$. In order to know if the next 1-bit is contained by $X_j$, Line 7 computes the position ($s$), where is the first 1-bit after the correspondent position of $i$ in $M$ ($beg_q+\Delta_i$). If $s$ is in the range of $X_j$ in $M$, $[beg_q, end_q]$, the next 1-bit is in $X_j$. In particular, that 1-bit is at distance $\Delta_{s}$ from the first position of $X_j$, hence the solution is $beg_{j}+\Delta_{s}$. 

\begin{figure}[t]
	\begin{minipage}[t]{.525\textwidth}
		\begin{algorithm}[H]
			\caption{{\bf succ}({$B$, $i$})\label{alg:successor}}
			$j \gets \lceil \frac{i}{\beta} \rceil$\\
			$q \gets rank_0(U,j)$ \tcp{mixed blocks}
			\If{$U[j]$}{
				\If{$O[j]$}{
					\Return $i$
				}
			}
			\Else{
				$s \gets succ(M, beg_{q} + \Delta_i)$\\
				\If(){$s \leq end_q$}{ 
					\Return $beg_{j} + \Delta_s$
				}
			}
			\Return  $jump(j,q)$
		\end{algorithm}
	\end{minipage}%
	\hfill\vrule\hfill
	\begin{minipage}[t]{.465\textwidth}
		\begin{algorithm}[H]
			\caption{{\bf jump}({$j$,$q$})\label{alg:jump}}
			$j' = succ(O, j+1)$\\
			\If{$U[j']$}{
				\Return $beg_{j'}$
			}
			\Else{
				$s' \gets succ(M, beg_{(q+1)})$\\
				\Return $beg_{j'}+ \Delta_{s'}$
			}
			\vspace{2.2cm}
		\end{algorithm}
	\end{minipage}
\end{figure}

Otherwise, the next 1-bit is not in $X_j$ because it is empty or the last 1-bit in $X_j$ is previous to $i$. The algorithm jumps to the next block with 1s by performing Algorithm~\ref{alg:jump}. Firstly, the next partition ($X_{j'}$) that contains at least one 1-bit is computed. If it is part of $\mathcal{O}$, Line 3, that block is full of 1s, and the solution is its first position. Otherwise, the result is located inside the $(q+1)$-th mixed block ($X_{j'}$) at position $\Delta_{s'}$, thus the algorithm returns $beg_{j'}+ \Delta_{s'}$.

Every $succ$ in plain bitvectors requires $o(n)$ extra space to solve them in $O(1)$ time. Consequently, \textit{zombit-vector} can compute \textit{succ} queries in $O(1)$ time with an extra-space of $o(n)$ bits. 
%The predecessor problem can be solved analogously, exchanging the successor operations on the plain bitvectors for predecessor. 

Therefore \textit{access}, \textit{rank}, and \textit{succ} can be solved in $O(1)$ time by using $O(k\beta + \frac{n}{\beta}) + o(n)$ bits. With the optimal value of $\beta$, $\sqrt{\frac{n}{k}}$, we can reduce the space to $O(\sqrt{kn})$ bits, and keep the last operations in $O(1)$ time. Furthermore, if we apply the \textit{zombit-vector} over $M$ recursively up to $c$ levels, we need to store $O(k^{1-\epsilon} n^{\epsilon})$ bits, where $\epsilon=\frac{1}{2^c}$. This recursive variant converges to $O(k)$ bits, and each operation can be solved in $O(\log\frac{n}{k})$ time. Recall that we do not discuss $pred(B,i)$ because it is symmetrical to $succ(B,i)$ and achieves identical theoretical bounds.

\Section{Experimental Evaluation}

\begin{figure}[!h]
	\centering    
	\includegraphics[width=0.91\textwidth]{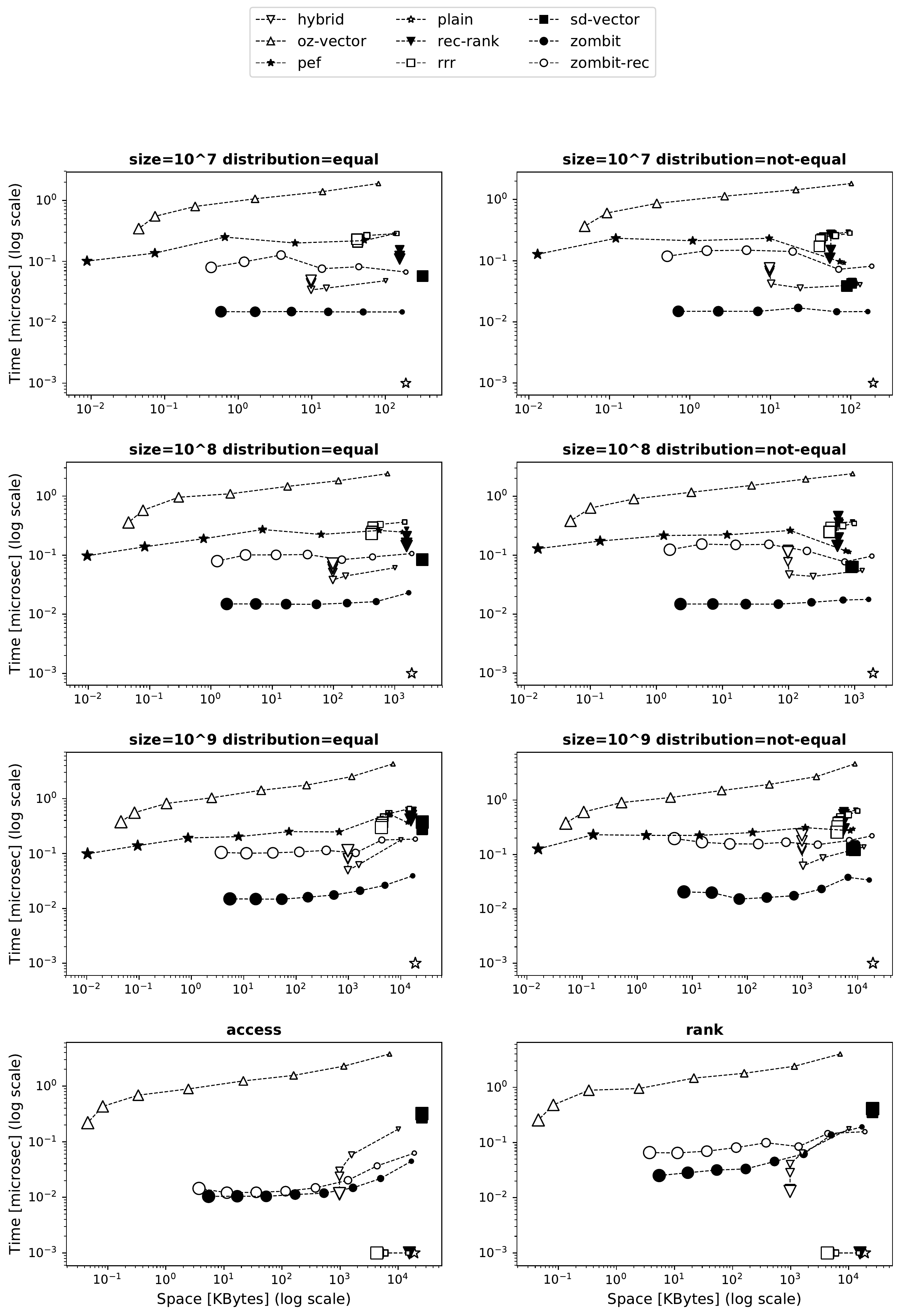}
	\caption{First six plots show the space and average response times on \textit{succ} queries for different sizes and distributions of runs. The last charts show the average response times of \textit{access} and \textit{rank} operations on a bitvector of size $10^9$ with an \textit{equal distribution} of runs. }
	\label{fig:times}
\end{figure}

	\textit{zombit-vector} was coded in C++,  using several data strucures from the SDSL library \cite{gbmp2014sea}. We have two implementations, the basic one, \textit{zombit}, and the recursive variant, \textit{zombit-rec}. Both implementations compute $\beta = \sqrt{\frac{n}{k}}$, and were compared with the different proposals presented in \textit{Background}  (\textit{plain}, \textit{rrr} \cite{raman2002succinct}, \textit{rec-rank} \cite{okanohara2007practical}, \textit{sd-array} \cite{okanohara2007practical}, \textit{oz-vector} \cite{Nav16}, and \textit{hybrid-vector} \cite{karkkainen2014hybrid}), and Partitioned Elias-Fano \cite{ottaviano2014partitioned} (\textit{pef}), a technique largely used in intersection of lists. Notice that the last baseline does not support \textit{rank} and \textit{access} operations on bitvectors, but it gets a good compression ratio/query time trade-off in \textit{succ} queries.
	
	Taking as reference the experimental evaluation of \cite{okanohara2007practical}, we built all the techniques over  bitvectors of sizes  $10^7$, $10^8$, and $10^9$ bits with several configurations. Let us denote with $run_0$ (resp. $run_1$) the mean length of runs of 0s (resp. 1s), inside the input bitvector. For each dimension of bitvector, we have two distributions of runs: \textit{equal distribution}, $run_1 \simeq run_0$, and \textit{not-equal distribution}, where $run_1 \simeq \frac{run_0}{8}$. 
	In each distribution we performed various executions, with different settings for $run_0$ and $run_1$. Given the $e$-th execution, $run_0=10^e$, and the values of $run_1$ are computed according to the chosen distribution. Notice that, in Figure~\ref{fig:times}, we can discern which value corresponds with each execution because the size of the markers increases while $run_0$ grows.

Over these data, we performed $1,000,000$ random successor queries and measured the average response time per operation. In most of the baseline cases, \textit{succ} was computed by using \textit{rank} and \textit{select}. However, some structures can solve successor more efficiently without using \textit{rank} and \textit{select}, specifically, \textit{sd-array} and \textit{hybrid-vector}. For a better comparison, in the last two structures, we have run the most efficient algorithms. The experiments were conducted on an Intel\textsuperscript{\textregistered} Xeon\textsuperscript{\textregistered} E5-2470 CPU @ 2.30GHz (32 cores) with 20MB of cache and 64 GB of RAM, running Debian GNU/Linux 10 with kernel 4.19.0-5 (64 bits), gcc version 8.3.0 with \texttt{-O3}.

\SubSection{Compression}

In terms of compression, if we compare the results between the two types of distributions, there is no significant difference. The behavior is very similar except for those techniques focused on sparse bitvectors, which obtain better compression in \textit{not-equal distribution}. The clear winner is \textit{pef}, but it is limited in functionality. It is followed by the \textit{oz-vector}, which needs $1.22\%-48.95\%$ of the space of \textit{zombit-rec}. Concerning the rest of the techniques, when $run_0$ is small, the \textit{hybrid-vector} is very competitive, for instance, it obtains the best compression when $run_0$ is lower than $10,000$ in the bitvector of size $10^9$. However, when $run_0$ grows, the size of the \textit{hybrid-vector} keeps constant, and it is improved by \textit{zombit}, which requires $0.72\%-54.02\%$ of its space. Besides, we can observe that there is a slight difference between our proposal and its recursive modification, in particular, \textit{zombit-rec} occupies $68.25\%-87.06\%$ of the space of \textit{zombit}.

\SubSection{Time performance}

As it is shown in the first six plots of Figure~\ref{fig:times}, the main competitor of our proposal for \textit{succ} operations is \textit{hybrid-vector}. However, it is beaten by \textit{zombit}, which becomes $3-12$ times faster and keeps those times constant. Our recursive variant, which slightly improves the space,  turns out $10\%-165\%$ slower than \textit{hybrid-vector}. Those results are similar to those obtained by \textit{rrr}, \textit{sd-array}, \textit{rec-rank} and \textit{pef}. The most time-consuming structure is \textit{oz-vector}. Though it can solve \textit{succ} with the same theoretical bound of \textit{hybrid-vector}, $O(\log n)$ time, in practice, it requires more binary searches, and turns $5-40$ times slower than \textit{hybrid-vector}. 

On the bottom of Figure~\ref{fig:times}, we compare the response times of \textit{access} and \textit{rank} in bitvectors with size $10^9$ and \textit{equal distribution} of runs. We can observe as \textit{zombit} and \textit{zombit-rec} are competitive in both operations, being close to the response times of \textit{hybrid-vector}. Therefore, \textit{zombit} becomes the structure with the best times in \textit{succ} queries, and keeps competitive times in \textit{access} and \textit{rank}.

\Section{Conclusions and Future Work}

We have proposed a structure, \textit{zombit}, which compresses bitvectors with large runs and can solve \textit{access}, \textit{rank} and \textit{successor/predecessor} queries in $O(1)$ time. We obtained a compression ratio of $0.04\%-26.33\%$, when the length of runs is larger than 100, and we can handle successor queries $3-12$ times faster than our immediate competitor.  Consequently, \textit{zombit} gets a good trade-off in terms of space and time on bitvectors with runs. 
A variant of our structure to obtain better compression was introduced, but in practice, it is $5-12$ times slower than \textit{zombit}, and it reduces to $68.25\%-87.06\%$ the space of our first proposal.

As future work, since we do not beat the space of \textit{hybrid-vector} in shorter runs, we will focus on improving the compression in that scenario. We plan to solve \textit{select} operations on \textit{zombit} efficiently by using $o(n)$ extra-space. Also, we will explore other areas where the successor and predecessor problem is relevant.

\Section{References}
\bibliographystyle{IEEEtran}
\bibliography{refs}

\end{document}